\documentclass{elsarticle-modified}

\usepackage[utf8]{inputenc}
\usepackage[T1]{fontenc}
\usepackage{fontaxes}
\usepackage{microtype}
\usepackage{mathrsfs}
\usepackage{mathtools}
\usepackage{todonotes}
\usepackage{xspace}
\usepackage{xcolor}
\usepackage{esvect}
\usepackage{xpatch}
\usepackage[all,defaultlines=3]{nowidow}
\usepackage[commandnameprefix=ifneeded]{changes}
\usepackage{amsthm}
\usepackage{amsmath}
\usepackage{amssymb}
\usepackage{cleveref}

\newtheorem{theorem}{Theorem}

\newtheorem{claim}[theorem]{Claim}
\newenvironment{claimproof}[1][\proofname]{%
  \begin{proof}[#1]%
}{%
  \end{proof}%
}

\usetikzlibrary{calc}
\usetikzlibrary{shapes}
\definecolor{darkgreen}{rgb}{10,117,28}
\usepackage{diagbox}
\usepackage{stackengine}
\usepackage{mathrsfs}

\definecolor{blue}{rgb}{0.1,0.2,0.5}
\definecolor{brown}{rgb}{0.6,0.6,0.2}

\makeatletter
\def\ifenv#1{
   \def\@tempa{#1}%
   \ifx\@tempa\@currenvir
      \expandafter\@firstoftwo
    \else
      \expandafter\@secondoftwo
   \fi
}
\let\wfs@comment@comment\comment
\let\comment\@undefined
\newcommand{\untoto}{\let\toto\@undefined}
\usepackage{changes}
\let\wfs@changes@comment\comment
\let\comment\@undefined

\newcommand\comment{%
    \ifthenelse{\equal{\@currenvir}{comment}}
    {\wfs@comment@comment}
    {\wfs@changes@comment}%
}
\makeatother

\newcommand{\rk}{\mathrm{rk}}

\newcommand{\Cc}{\mathscr{C}}

\renewcommand{\phi}{\varphi}

\newcommand{\x}{\bar{x}}

\newcommand{\N}{\mathbb{N}}

\usepackage{accents}

\usepackage{wasysym}
\usepackage{xspace}

\title{A Note on Constructive Canonical Splitter Strategies\\ in Nowhere Dense Graph Classes}
\author[inst1]{Janne Fuchser}
\ead{jfuchser@uni-bremen.de}
\author[inst2]{Nikolas Mählmann}
\ead{maehlmann@mimuw.edu.pl}
\author[inst1]{Sebastian Siebertz}
\ead{siebertz@uni-bremen.de}
\address[inst1]{University of Bremen, Germany}
\address[inst2]{University of Warsaw, Poland}

\begin{document}

\begin{frontmatter}
\begin{abstract}
The radius-$r$ splitter game is played on a graph $G$
between two players: Splitter and Connector. 
In each round, Connector selects a vertex $v$, and the current game arena is restricted to the radius-$r$ neighborhood of $v$. 
Then Splitter removes a vertex from this restricted subgraph. 
The game ends, and Splitter wins, when the arena becomes empty. 
Splitter aims to end the game as quickly as possible, while Connector tries to prolong it for as long as possible.
The splitter game was introduced by Grohe, Kreutzer and Siebertz to characterize nowhere dense graph classes. 
They showed that a class $\Cc$ of graphs is nowhere dense if and only if for every radius $r$ there exists a number $k$ such that Splitter has a strategy on every $G\in \Cc$ to win the radius-$r$ splitter game in at most $k$ rounds. 
It was recently proved by Ohlmann et al.\ that for every nowhere dense class $\Cc$ and every radius $r$ there are only a bounded number of possible Splitter moves that are progressing, that is, moves that lead to an arena where Splitter can win in one less round.
The proof of Ohlmann et al.\ is based on the compactness theorem and does not give a constructive bound on the number of progressing moves. 
In this work, we give a simple constructive proof, showing that if Splitter can force a win in the radius-$r$ game in $k$ rounds, then there are at most $(2r+1)^{\,2^{k-1}-1}$ progressing moves.
\end{abstract}
\begin{keyword}
splitter game, nowhere dense graph classes
\end{keyword}
\end{frontmatter}


Nowhere dense graph classes were introduced by Ne\v{s}et\v{r}il and Ossona de Mendez~\cite{nevsetvril2011nowhere}, and provide a robust framework for capturing uniform sparsity in graphs.\footnote{All graphs in this paper are finite, undirected and simple. For a graph $G$ and $A\subseteq V(G)$ we write $G[A]$ for the subgraph of $G$ induced by $A$. For a vertex $v\in V(G)$ and $r\in \N$ we write~$N^G_r[v]$ for the closed $r$-neighborhood of $v$, that is, the set of all vertices at distance at most $r$ from $v$ in $G$. If $G$ is clear from the context, we omit the superscript.
In the following, $r$ will always denote a positive integer. We write $G-v$ for the graph induced by $V(G)\setminus\{v\}$. }
Formally, a graph class $\Cc$ is \emph{nowhere dense} if for every positive integer $r$ there exists a positive integer $t = t(r)$ such that the $r$-subdivision of the complete graph $K_t$ is not a subgraph of any graph in $\Cc$.
Nowhere denseness is a natural generalization of other notions of sparsity, including bounded degree, planarity, and excluded minors or topological minors. 
For example, if $G$ excludes a graph $H$ as a minor, then it excludes $H$ as a topological minor, and writing $t=|V(H)|$, in particular the $r$-subdivision of $K_t$ for every positive integer $r$. 

As one of the main tools to efficiently solve the first-order model checking problem on nowhere dense graph classes, Grohe, Kreutzer, and Siebertz in~\cite{GroheKS17} introduced a very intuitive characterization of nowhere dense classes in terms of a game, called the \emph{splitter game}\footnote{We consider here what is called the \emph{simple splitter game} in~\cite{GroheKS17}. In the general splitter game, Splitter is allowed to delete a constant number of vertices in every round at once, which however is equivalent to the simple splitter game by Remark 4.4.\ of~\cite{GroheKS17}.}. 
The \emph{radius-$r$ splitter game} is played on a graph $G$ between two players: \emph{Splitter} and \emph{Connector}. The game starts with the arena $G_1=G$ and proceeds in rounds $i=1,2,\ldots$, where in each round 
\begin{itemize}
    \item Connector chooses a vertex $c_i \in V(G_i)$, and then 
    \item Splitter chooses a vertex $s_i \in N_r[c_i]$, and the new arena is defined as
    \[
    G_{i+1} = G_i[N_r[c_i]] - s_i, 
    \]
    where $N_r[c_i]$ is the $r$-neighborhood of $c_i$ in $G_i$.
\end{itemize}

Splitter wins and the game ends when $G_{i+1}$ is the empty graph, otherwise the game continues with $G_{i+1}$.
Splitter's goal is to win as quickly as possible, while Connector aims to prolong the game as long as possible. 
We define the \emph{$r$-splitter rank} of~$G$, denoted $\rk_r(G)$, as the least number of rounds in which Splitter can force a win in the radius-$r$ splitter game on $G$.
Note that for any subgraph $H\subseteq G$, it holds that $\rk_r(H)\leq\rk_r(G)$.
Formally, we have $\rk_r(K_1) = 1$ and
\[
    \rk_r(G) = 1+ \max_{c \in V(G)} \min_{s \in N_r[c]} \rk_r \big(G[N_r[c]] - s \big).
  \]


\begin{theorem}[Grohe et al.~\cite{GroheKS17}]\label{thm:game-characterization}
A graph class $\Cc$ is nowhere dense if and only if for every radius $r$, there exists $k(r)$ such that $\rk_r(G)\leq k(r)$ for all $G\in \Cc$.
\end{theorem}

  

Besides the number of rounds needed to win, it is of course of greatest interest to know which moves lead Splitter to a fastest win. 
Given a Connector move $c \in V(G)$, we call a Splitter move $s \in V(G)$ \emph{$r$-progressing against $c$} if the $r$-splitter rank of $G[N_r[c]] - s$ is strictly smaller than the $r$-splitter rank of $G[N_r[c]]$.
In particular any move that is $r$-progressing against $c$ must be contained in $N_r[c]$.

It was recently proved by Ohlmann et al.~\cite{OhlmannPPT23} that for every nowhere dense class and every positive integer $r$, for every Connector move $c$ in a graph $G\in \Cc$, there are only a bounded number of possible Splitter moves that are progressing against $c$, where the upper bound depends on the radius~$r$.

\begin{theorem}[Ohlmann et al.~\cite{OhlmannPPT23}]\label{thm:progressive}
Let $\Cc$ be a nowhere dense graph class. 
For every radius $r$ there exists $f=f(r)$ such that for every graph $G \in \Cc$, and every Connector move $c$, there are at most $f$ many $r$-progressing moves against $c$ in $G$.
\end{theorem}

Note that being a progressive move is a first-order definable property, so that the model checking algorithm of Grohe et al.\ can be used to efficiently find all of these progressive moves. 
This first-order definability of progressing moves also led Ohlmann et al.\ to their proof that there are only finitely many progressing moves, based on the compactness theorem for first-order logic. 
The proof by compactness is very elegant and short but on the downside does not yield a constructive bound on the number of progressing moves. 

In this short note we give a simple constructive proof, showing that if the $r$-splitter rank of $G$ is $k$, then for every Connector move $c$ there are at most $f(k)=(2r+1)^{\,2^{k-1}-1}$ $r$-progressing moves against $c$.
Hence by Theorem~\ref{thm:game-characterization}, $f(k(r))=(2r+1)^{2^{k(r)-1}-1}$ is a constructive upper bound for $f$ from Theorem~\ref{thm:progressive}. 

\begin{theorem}\label{thm:main}
Fix a radius $r$ and let $G$ be a graph with $r$-splitter rank $k$. 
Then there are at most $f(k)=(2r+1)^{\,2^{k-1}-1}$ $r$-progressing Splitter moves against any Connector move $c$ in $G$.
\end{theorem}

Apart from providing concrete, computable values for $f(k)$, \Cref{thm:main} also differs from the statement of \Cref{thm:progressive} by working in the (more general) setting of arbitrary graphs, instead of requiring to fix a nowhere dense graph class.
We note that the non-constructive proof of \Cref{thm:progressive} given in \cite{OhlmannPPT23} can easily be modified to also work in this more general setting.

We prove the following strengthening of \Cref{thm:main}.

\begin{theorem}\label{thm:constructive-splitter-new}
Fix a radius $r$ and let $G$ be a graph of $r$-splitter rank $k$. 
There exists an induced subgraph $H\subseteq G$ with $|H|\leq f(k)$ that has $r$-splitter rank~$k$. 
\end{theorem}

It implies \Cref{thm:main} as follows.

\begin{proof}[Proof of \Cref{thm:main} using \Cref{thm:constructive-splitter-new}]
    Let $G$ be a graph of $r$-splitter rank $k$ and~$c$ be a Connector move in~$G$.
    The graph $G_c := G[N_r[c]]$ has $r$-splitter rank $k_c \leq k$. By \Cref{thm:constructive-splitter-new}, it contains an induced subgraph $H_c \subseteq G_c$ of size $|H_c| \leq f(k_c) \leq f(k)$.
    Consider now a Splitter move $s \in V(G_c) \setminus V(H_c)$. As $G_c - s$ still contains $H_c$, we have 
    \[
        \rk_r(G_c -s) = \rk_r(H_c) = \rk_r(G_c) = k_c,    
    \]
    which means that $s$ is not $r$-progressing against $c$.
    Hence, all $r$-progressing moves against $c$ are contained in $H_c$, which implies that their number is bounded by~$f(k)$.    
\end{proof}

It therefore only remains to prove \Cref{thm:constructive-splitter-new}.
Before we prove it constructively, we will present another short and elegant non-constructive, model-theoretic proof (yielding finite but non-constructive bounds for $f$), which in fact was our initial inspiration for the statement.

\begin{proof}[Sketch of a non-constructive proof of \Cref{thm:constructive-splitter-new}]
    It is easy to verify that for every fixed $r$ and $k$, there is a first-order sentence $\sigma_{r,k}$ such that for every graph~$G$
    \[
        G \models \sigma_{r,k} \iff \rk_r(G) \geq k.
    \]

    Informally, $\sigma_{r,k}$ expresses that Connector can survive for $k$ rounds in the radius-$r$ splitter game.
    More precisely, it encodes the alternating condition
    \[
        \exists x_1 \;\forall y_1 \;\exists x_2 \;\forall y_2 \;\cdots\; \exists x_k \;\forall y_k
    \]
    where each Splitter move $y_i$ must lie in the radius-$r$ neighborhood of the current Connector move, and each Connector move $x_{i+1}$ is chosen in the remaining arena after deleting the previously chosen Splitter vertices.
    Thus, $\sigma_{r,k}$ states that Connector has a strategy to keep the play alive for $k$ rounds.
    Since $r$ is fixed, the relation ``$u$ lies in the radius-$r$ neighborhood of $v$'' is first-order definable (via a bounded-length path formula), and hence all locality constraints can be expressed in first-order logic.

    We refrain from spelling out the full formula $\sigma_{r,k}$ explicitly, as it is cumbersome and notationally heavy, while its structure is exactly as described above.

    By definition of the splitter rank, this formula is preserved by induced supergraphs: if a graph $G$ satisfies $\sigma_{r,k}$, then every graph $H$ that contains $G$ as an induced subgraph also satisfies $\sigma_{r,k}$.
    By the \L{}o\'s-Tarski theorem (see, e.g., \mbox{\cite[Cor.\ 6.5.5]{hodges_1993}}), $\sigma_{r,k}$ is equivalent to an existential formula. 
    Note that the theorem is non-constructive and we do not know the existential formula.
    By the structure of existential formulas, there must exist a finite list of finite graphs $\cal H$ such that
    every graph in $\cal H$ has $r$-splitter rank $k$, and
    every graph with $r$-splitter rank $k$ contains a graph from $\cal H$ as an induced subgraph.
    Hence, we can bound $f(k)$ by the size of the largest graph in $\cal H$.
\end{proof}

It is finally time to give our constructive proof of \Cref{thm:constructive-splitter-new}.

\begin{proof}[Proof of \Cref{thm:constructive-splitter-new}]
We proceed by induction on the $r$-splitter rank $k$ of $G$.
For $k=1$ we have $H=G=K_1$, which defines $f(1):= 1$ as the initial value for the function~$f$.

For the inductive step, assume the statement holds for $r$-splitter rank $k$ and let $G$ be a graph with $\rk_r(G)=k+1$. 
Let $B$ be a minimal induced subgraph of $G$ with rank $k+1$, i.e., every proper induced subgraph of $B$ has rank $\leq k$. 
Such a minimal $B$ exists, although it isn't necessarily unique.
We establish three properties of $B$.

    \begin{claim}\label{clm:center-of-B}
        There exists a central vertex $c_1\in V(B)$, i.e., $B = B[N_r^B[c_1]]$.
    \end{claim}
    \begin{claimproof}
        As $\rk_r(B)=k+1$ there exists an optimal Connector move $c_1\in V(B)$ that Connector can make so that the game lasts for $k+1$ rounds. 
        In particular, $B[N_r^B[c_1]]$ is a subgraph of $B$ with $r$-splitter rank $k+1$. 
        By minimality of $B$, this cannot be a proper subgraph of $B$, so $B=B[N_r^B[c_1]]$.
    \end{claimproof}

    \begin{claim}\label{clm:all-progressing-in-B}
        Every vertex of $B$ is \emph{progressing in $B$}, that is, for every vertex $s\in V(B)$ we have $\rk_r(B-s)=k$.
    \end{claim}
    \begin{claimproof}
        By minimality of $B$ we have 
        $\rk_r(B-s)\leq k$. 
        Moreover, \mbox{$\rk_r(B-s)< k$} is impossible, because then we would have $\rk_r(B)\leq\rk_r(B-s)+1<k+1$, in contradiction to $\rk_r(B)=k+1$.
    \end{claimproof}

    From this property of $B$ and our induction hypothesis we derive the following.

    \begin{claim}\label{clm:use-induction-hypothesis}
        For every vertex $s\in V(B)$, there exists an induced subgraph $H_s\subseteq B-s$ such that $\rk_r(H_s)=k$ and $|H_s|\leq f(k)$.
    \end{claim}

    Using Claim~\ref{clm:use-induction-hypothesis} and Claim~\ref{clm:center-of-B}, we can construct our small subgraph~$H$ of~$B$ with $\rk_r(H)=k+1$:
    Choose any $s\in B$ and let $H_s$ be as in Claim~\ref{clm:use-induction-hypothesis}.
    Let~$H$ be the subgraph induced by the following vertices: 
    \begin{itemize}
        \item all vertices of $H_s$,
        \item all vertices of $H_v$ for every $v\in V(H_s)$, and
        \item the vertices of a path of length at most $r$ from the center $c_1$ of $B$ to each vertex in $H_s$ and the vertices of paths from $c_1$ to every vertex in every~$H_v$. (This center $c_1$ and these paths exist by Claim~\ref{clm:center-of-B}.)
        Denote the set of vertices of these paths by $P$ and the set of inner vertices of these paths by~$Q$. 
    \end{itemize}
    
Note that $H$ includes $c_1$, as it is the starting point of all the paths we include in~$H$, even though possibly $H_s$ and all $H_v$ do not include it. See \Cref{fig:obstruction} for an illustration.

\begin{figure}[htbp]
    \centering
    \includegraphics{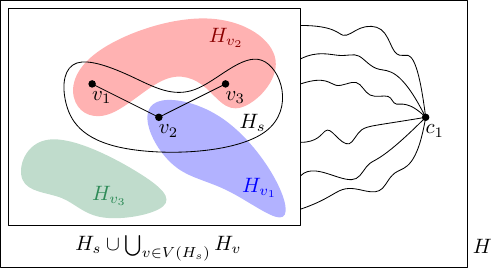}
    \caption{Construction of the graph $H$. Each $H_{v}$ avoids $v$.}
    \label{fig:obstruction}
\end{figure}

\smallskip
First, we observe that $H$ has $\rk_r(H) = k+1$. 
\begin{itemize}
    \item As a subgraph of $B$, it has $\rk_r(H) \leq \rk_r(B) = k+1$.
    \item Connector can start by playing $c_1$ such that $H[N_r^H[c_1]]=H$. 
    If Splitter then deletes a vertex $v\in V(H_s)$, then the subgraph $H_v$ of rank $k$ survives. 
    Otherwise, the subgraph $H_s$ of rank $k$ survives. 
    In both cases, the game takes $k+1$ rounds, giving us $\rk_r(H) \geq k+1$.
  \end{itemize}
Second, we bound $|H|$ as follows. 
\begin{align*}
  |H| & = |H_s\cup\bigcup_{v\in H_s} H_v \cup P|\\
  & \leq |H_s| + \left(\sum_{v\in H_s}|H_v|\right) + |Q| + |\{c_1\}| \\
  & \leq f(k) + f(k)^2 + (f(k) + f(k)^2) \cdot (r-1) + 1 \\
  &  = (f(k) + f(k)^2) \cdot r +1 \\
  &= r f(k)^2 + r f(k) + 1\\
  & \eqqcolon f(k+1)\, ,
\end{align*}
where we obtain the third from the second line by using that $\sum_{v\in H_s}|H_v|$ contains at most $f(k)$ many summands, each of which is $\leq f(k)$, and that $Q$ contains at most $r-1$ many inner path vertices for any vertex in $H_s\cup\bigcup_{v\in H_s} H_v$.

We have $f(k+1)\leq r f(k)^2 + (r+1) f(k)^2=(2r+1)\, f(k)^2$, since $f(k)\geq 1$, and define 
the sequence $g$ by $g(1) = f(1) = 1$, and $g(k+1) = (2r+1)\, g(k)^2$. 
By induction, we obtain $f(k) \le g(k)$ for all $k\ge 1$. 
The recurrence for $g$ solves explicitly to
$g(k) = (2r+1)^{\,2^{k-1}-1}$, and hence 
\[
f(k) \;\le\; (2r+1)^{\,2^{k-1}-1}.
\]

  This finishes our proof that indeed, there exists a small induced subgraph $H$ of $B$ (which is in turn an induced subgraph of $G$) that also has $\rk_r(H)=k+1$, so the induction step is complete.
\end{proof}

\bibliographystyle{plain}
\bibliography{ref}

\end{document}